\pdfoutput=1

\documentclass[12pt,a4paper]{article}

\usepackage{ifthen} 
\newboolean{pdflatex}
\setboolean{pdflatex}{true} 

\newboolean{articletitles}
\setboolean{articletitles}{true} 

\newboolean{uprightparticles}
\setboolean{uprightparticles}{false} 

\newboolean{inbibliography}
\setboolean{inbibliography}{false} 

\newboolean{forprl}
\setboolean{forprl}{false} 

\def\paperauthors{LHCb collaboration} 
\def\paperasciititle{chic1 and chi2 resonance parameters with the decays chicc1,chic2->J/psi mumu} 
\def\papertitle{$\chicone$ and $\chictwo$~resonance parameters with the~decays $\chi_{c1,c2}\to\jpsi\mu^+\mu^-$} 
\def\paperkeywords{{High Energy Physics}, {LHCb}} 
\def\papercopyright{CERN on behalf of the LHCb collaboration}
\def\paperlicence{CC-BY-4.0}
\def\paperlicenceurl{https://creativecommons.org/licenses/by/4.0/}

\usepackage[top=1in, bottom=1.25in, left=1in, right=1in]{geometry}

\usepackage{microtype}
\usepackage{lineno}  
\usepackage{xspace} 
\usepackage{caption} 

\usepackage{graphicx}  
\usepackage{color}
\usepackage{colortbl}
\graphicspath{{./figs/}} 

\usepackage{amsmath} 
\usepackage{amssymb}
\usepackage{amsfonts}
\usepackage{upgreek} 

\newcommand*\patchAmsMathEnvironmentForLineno[1]{%
\expandafter\let\csname old#1\expandafter\endcsname\csname #1\endcsname
\expandafter\let\csname oldend#1\expandafter\endcsname\csname
end#1\endcsname
 \renewenvironment{#1}%
   {\linenomath\csname old#1\endcsname}%
   {\csname oldend#1\endcsname\endlinenomath}%
}
\newcommand*\patchBothAmsMathEnvironmentsForLineno[1]{%
  \patchAmsMathEnvironmentForLineno{#1}%
  \patchAmsMathEnvironmentForLineno{#1*}%
}
\AtBeginDocument{%
\patchBothAmsMathEnvironmentsForLineno{equation}%
\patchBothAmsMathEnvironmentsForLineno{align}%
\patchBothAmsMathEnvironmentsForLineno{flalign}%
\patchBothAmsMathEnvironmentsForLineno{alignat}%
\patchBothAmsMathEnvironmentsForLineno{gather}%
\patchBothAmsMathEnvironmentsForLineno{multline}%
\patchBothAmsMathEnvironmentsForLineno{eqnarray}%
}

\usepackage{hyperxmp}

\usepackage[pdftex,
            pdfauthor={\paperauthors},
            pdftitle={\paperasciititle},
            pdfkeywords={\paperkeywords},
            pdfcopyright={Copyright (C) \papercopyright},
            pdflicenseurl={\paperlicenceurl}]{hyperref}

\usepackage[all]{hypcap} 

\usepackage{xspace} 
\usepackage{upgreek}

\def\MagUp {\mbox{\em Mag\kern -0.05em Up}\xspace}

\ifthenelse{\boolean{uprightparticles}}%
{

 \def\Pmu         {\ensuremath{\upmu}\xspace}

 \def\Pchi        {\ensuremath{\upchi}\xspace}                 
 \def\Ppsi        {\ensuremath{\uppsi}\xspace}

 \def\PDelta      {\ensuremath{\Delta}\xspace}                 
 \def\PXi      {\ensuremath{\Xi}\xspace}                 
 \def\PLambda      {\ensuremath{\Lambda}\xspace}                 
 \def\PSigma      {\ensuremath{\Sigma}\xspace}                 
 \def\POmega      {\ensuremath{\Omega}\xspace}                 
 \def\PUpsilon      {\ensuremath{\Upsilon}\xspace}                 
 
 \def\PB      {\ensuremath{\mathrm{B}}\xspace}                 
                  
 \def\PD      {\ensuremath{\mathrm{D}}\xspace}

 \def\PJ      {\ensuremath{\mathrm{J}}\xspace}                 
 \def\PK      {\ensuremath{\mathrm{K}}\xspace}

 \def\Pb      {\ensuremath{\mathrm{b}}\xspace}                 
 \def\Pc      {\ensuremath{\mathrm{c}}\xspace}

 \def\Pi      {\ensuremath{\mathrm{i}}\xspace}

}
{

 \def\Pmu         {\ensuremath{\mu}\xspace}

 \def\Pchi        {\ensuremath{\chi}\xspace}                 
 \def\Ppsi        {\ensuremath{\psi}\xspace}                 
                  
 \mathchardef\PDelta="7101
 \mathchardef\PXi="7104
 \mathchardef\PLambda="7103
 \mathchardef\PSigma="7106
 \mathchardef\POmega="710A
 \mathchardef\PUpsilon="7107
                  
 \def\PB      {\ensuremath{B}\xspace}                 
                  
 \def\PD      {\ensuremath{D}\xspace}

 \def\PJ      {\ensuremath{J}\xspace}                 
 \def\PK      {\ensuremath{K}\xspace}

 \def\Pb      {\ensuremath{b}\xspace}                 
 \def\Pc      {\ensuremath{c}\xspace}

 \def\Pi      {\ensuremath{i}\xspace}

}

\makeatletter
\ifcase \@ptsize \relax
  \newcommand{\miniscule}{\@setfontsize\miniscule{4}{5}}
\or
  \newcommand{\miniscule}{\@setfontsize\miniscule{5}{6}}
\or
  \newcommand{\miniscule}{\@setfontsize\miniscule{5}{6}}
\fi
\makeatother

\DeclareRobustCommand{\optbar}[1]{\shortstack{{\miniscule (\rule[.5ex]{1.25em}{.18mm})}
  \\ [-.7ex] $#1$}}

\def\mumu       {{\ensuremath{\Pmu^+\Pmu^-}}\xspace}

\def\cquark    {{\ensuremath{\Pc}}\xspace}

\def\bquark    {{\ensuremath{\Pb}}\xspace}

\def\kaon    {{\ensuremath{\PK}}\xspace}
  \def\Kbar    {{\kern 0.2em\overline{\kern -0.2em \PK}{}}\xspace}

\def\KorKbar    {\kern 0.18em\optbar{\kern -0.18em K}{}\xspace}

\def\KS      {{\ensuremath{\kaon^0_{\mathrm{ \scriptscriptstyle S}}}}\xspace}

  \def\Dbar    {{\kern 0.2em\overline{\kern -0.2em \PD}{}}\xspace}

\def\DorDbar    {\kern 0.18em\optbar{\kern -0.18em D}{}\xspace}

\def\B       {{\ensuremath{\PB}}\xspace}
\def\Bbar    {{\ensuremath{\kern 0.18em\overline{\kern -0.18em \PB}{}}}\xspace}

\def\BorBbar    {\kern 0.18em\optbar{\kern -0.18em B}{}\xspace}

\def\Bu      {{\ensuremath{\B^+}}\xspace}

\def\jpsi     {{\ensuremath{{\PJ\mskip -3mu/\mskip -2mu\Ppsi\mskip 2mu}}}\xspace}

\def\chicone  {{\ensuremath{\Pchi_{\cquark 1}}}\xspace}
\def\chictwo  {{\ensuremath{\Pchi_{\cquark 2}}}\xspace}
  \def\Y#1S{\ensuremath{\PUpsilon{(#1S)}}\xspace}

\def\Lbar        {{\ensuremath{\kern 0.1em\overline{\kern -0.1em\PLambda}}}\xspace}
\def\LorLbar    {\kern 0.18em\optbar{\kern -0.18em \PLambda}{}\xspace}

\def\to                 {\ensuremath{\rightarrow}\xspace}

\def\AT#1     {\ensuremath{A_{\mathrm{T}}^{#1}}\xspace}           

\def\C#1      {\ensuremath{\mathcal{C}_{#1}}\xspace}                       
\def\Cp#1     {\ensuremath{\mathcal{C}_{#1}^{'}}\xspace}                    
\def\Ceff#1   {\ensuremath{\mathcal{C}_{#1}^{\mathrm{(eff)}}}\xspace}        
\def\Cpeff#1  {\ensuremath{\mathcal{C}_{#1}^{'\mathrm{(eff)}}}\xspace}       
\def\Ope#1    {\ensuremath{\mathcal{O}_{#1}}\xspace}                       
\def\Opep#1   {\ensuremath{\mathcal{O}_{#1}^{'}}\xspace}                    

\newcommand{\tev}{\ifthenelse{\boolean{inbibliography}}{\ensuremath{~T\kern -0.05em eV}}{\ensuremath{\mathrm{\,Te\kern -0.1em V}}}\xspace}
\newcommand{\gev}{\ensuremath{\mathrm{\,Ge\kern -0.1em V}}\xspace}
\newcommand{\mev}{\ensuremath{\mathrm{\,Me\kern -0.1em V}}\xspace}
\newcommand{\kev}{\ensuremath{\mathrm{\,ke\kern -0.1em V}}\xspace}
\newcommand{\ev}{\ensuremath{\mathrm{\,e\kern -0.1em V}}\xspace}
\newcommand{\gevc}{\ensuremath{{\mathrm{\,Ge\kern -0.1em V\!/}c}}\xspace}
\newcommand{\mevc}{\ensuremath{{\mathrm{\,Me\kern -0.1em V\!/}c}}\xspace}
\newcommand{\gevcc}{\ensuremath{{\mathrm{\,Ge\kern -0.1em V\!/}c^2}}\xspace}
\newcommand{\gevgevcccc}{\ensuremath{{\mathrm{\,Ge\kern -0.1em V^2\!/}c^4}}\xspace}
\newcommand{\mevcc}{\ensuremath{{\mathrm{\,Me\kern -0.1em V\!/}c^2}}\xspace}

\def\invfb   {\ensuremath{\mbox{\,fb}^{-1}}\xspace}

\newcommand{\stat}{\ensuremath{\mathrm{\,(stat)}}\xspace}
\newcommand{\syst}{\ensuremath{\mathrm{\,(syst)}}\xspace}

\def\gsim{{~\raise.15em\hbox{$>$}\kern-.85em
          \lower.35em\hbox{$\sim$}~}\xspace}
\def\lsim{{~\raise.15em\hbox{$<$}\kern-.85em
          \lower.35em\hbox{$\sim$}~}\xspace}

\def\sPlot{\mbox{\em sPlot}\xspace}

\def\pt         {\mbox{$p_{\mathrm{ T}}$}\xspace}

\def\evtgen     {\mbox{\textsc{EvtGen}}\xspace}

\def\geant      {\mbox{\textsc{Geant4}}\xspace}

\def\photos     {\mbox{\textsc{Photos}}\xspace}

\def\pythia     {\mbox{\textsc{Pythia}}\xspace}

\def\tell1  {TELL1\xspace}
\def\ukl1   {UKL1\xspace}

\newcommand{\eg}{\mbox{\itshape e.g.}\xspace}

\usepackage{cite} 
\usepackage{mciteplus}

\usepackage{longtable} 
\usepackage{rotating}

\begin{document}

\renewcommand{\thefootnote}{\fnsymbol{footnote}}
\setcounter{footnote}{1}


\begin{titlepage}
\pagenumbering{roman}

\vspace*{-1.5cm}
\centerline{\large EUROPEAN ORGANIZATION FOR NUCLEAR RESEARCH (CERN)}
\vspace*{0.5cm}
\noindent
\begin{tabular*}{\linewidth}{lc@{\extracolsep{\fill}}r@{\extracolsep{0pt}}}
\ifthenelse{\boolean{pdflatex}}
{\vspace*{-1.5cm}\mbox{\!\!\!\includegraphics[width=.14\textwidth]{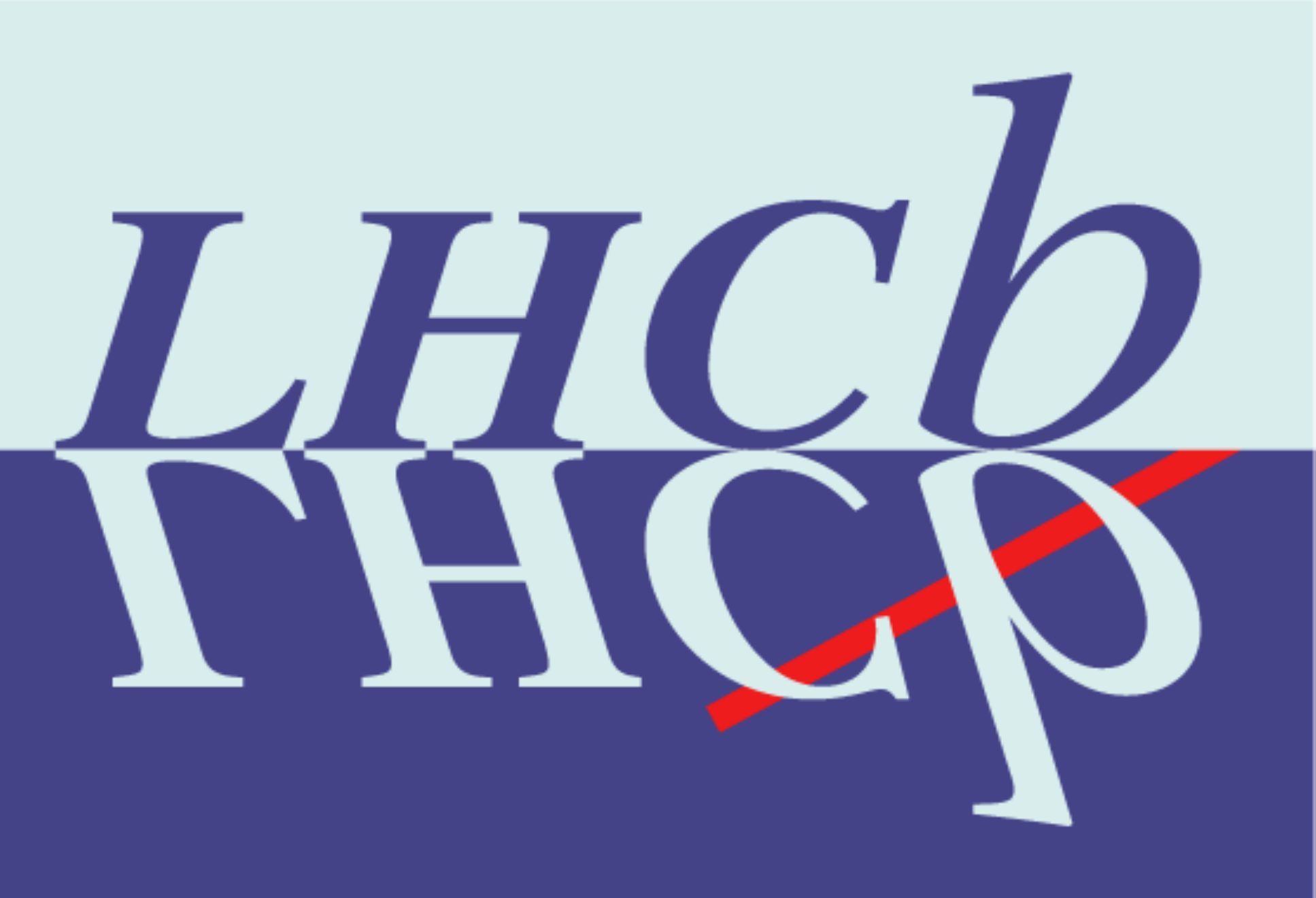}} & &}%
{\vspace*{-1.2cm}\mbox{\!\!\!\includegraphics[width=.12\textwidth]{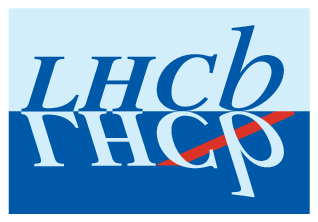}} & &}%
\\
 & & CERN-EP-2017-228 \\  
 & & LHCb-PAPER-2017-036 \\  
 & & September 12, 2017 \\ 
\end{tabular*}

\vspace*{2.0cm}

{\normalfont\bfseries\boldmath\huge
\begin{center}
  \papertitle 
\end{center}
}

\vspace*{1.0cm}

\begin{center}
\paperauthors\footnote{Authors are listed at the end of this Letter.}
\end{center}

\vspace{\fill}

\begin{abstract}
  \noindent
  The decays $\chicone \rightarrow \jpsi \mu^+ \mu^-$ and
  $\chictwo \rightarrow \jpsi \mu^+ \mu^-$ are observed and used to
  study the resonance parameters of the $\chicone$ and $\chictwo$
  mesons. The masses of these states are measured to be
  \begin{align*}
    m(\chicone) = 3510.71 \pm 0.04\stat  \pm 0.09\syst\mev\,, \\
    m(\chictwo) = 3556.10 \pm 0.06\stat  \pm 0.11\syst\mev\,, 
  \end{align*}
  where the knowledge of the momentum scale for charged particles
  dominates the systematic uncertainty. 
  The momentum\nobreakdash-scale uncertainties largely cancel in the mass difference
  \begin{equation*}
    m(\chictwo) - m(\chicone) = 45.39 \pm 0.07\stat \pm 0.03\syst\mev\,.  \\
  \end{equation*}
  The natural width of the $\chictwo$ meson is measured to be
  \begin{equation*}
    \Gamma(\chictwo) = 2.10 \pm 0.20\stat \pm 0.02\syst\mev\,. \\
  \end{equation*}
  These results are in good agreement with and have comparable precision
  to the current world averages.
\end{abstract}

\vspace*{1.0cm}

\begin{center}
  Published in Phys.~Rev.~Lett.~{\bf{119}}, 221901 (2017).
\end{center}

\vspace{\fill}

{\footnotesize 
\centerline{\copyright~\papercopyright, licence \href{\paperlicenceurl}{\paperlicence}.}}
\vspace*{2mm}

\end{titlepage}


\newpage
\setcounter{page}{2}
\mbox{~}
%
%
%
%

\cleardoublepage


\renewcommand{\thefootnote}{\arabic{footnote}}
\setcounter{footnote}{0}



\pagestyle{plain} 
\setcounter{page}{1}
\pagenumbering{arabic}


%


   
Studies of the properties and production of quarkonia at hadron
colliders provide an important testing ground for Quantum
Chromodynamics~\cite{Brambilla:2010cs}.
Measurements of the spectra test potential models~\cite{Eichten:1978tg}
whilst the production rate can be calculated pertubatively in
nonrelativistic effective field theories such
as NRQCD~\cite{Brambilla:2004jw}. Most studies of $\chicone$ and $\chictwo$
mesons at hadron colliders have exploited the radiative decays
$\chi_{c1,c2} \rightarrow \jpsi \gamma$  
with the subsequent decay $\jpsi \rightarrow \mu^+ \mu^-$~\cite{Abulencia:2007bra,LHCb-PAPER-2011-019,LHCb-PAPER-2013-028,Chatrchyan:2012ub,ATLAS:2014ala}. 
The branching fractions for these processes are large, allowing a signal
to be observed despite high background.
   
Recently, the BESIII collaboration~\cite{Ablikim:2017kia} 
reported the first observation of the electromagnetic Dalitz decays~\cite{Dalitz:1951aj}
of  $\chi_{c0}$, $\chi_{c1}$ and $\chi_{c2}$~mesons 
into the~$\jpsi e^+ e^-$~final state.
This Letter reports
the first observation of the  $\chicone
\rightarrow \jpsi \mu^+ \mu^-$ and $\chictwo
\rightarrow \jpsi \mu^+ \mu^-$ decay modes, 
using $\jpsi\to  \mu^+\mu^-$~decays.
These decays are used to measure the $\chicone$ and
$\chictwo$~masses together with the $\chictwo$ natural width. 
The event topology with four
muons in the final state provides a clean signature that is ideal
for studies in hadron collisions. 

This analysis uses the LHCb data set collected in $pp$ collisions up to the end of 2016.
The data collected at centre-of-mass energies of 7 and 8\tev corresponds to
integrated liminosities of 1 and 2\invfb and is collectively referred to as Run\,1,
while  data collected at centre-of-mass energy of 13\tev corresponds to
1.9\invfb and is referred to as Run\,2.


The LHCb detector is a single-arm spectrometer covering the pseudorapidity range 
$2 < \eta < 5$, described in detail in 
Refs.~\cite{Alves:2008zz,LHCb-DP-2014-002}. The detector includes a high-precision tracking system
consisting of a silicon-strip vertex detector~\cite{LHCb-DP-2014-001}, a large-area silicon-strip detector located
upstream of a dipole magnet with a bending power 
of about
4\,Tm, and three stations of silicon-strip detectors and straw
drift tubes~\cite{LHCb-DP-2013-003} placed downstream of the magnet.
The tracking system measures the momentum of charged particles with 
a relative uncertainty that varies from 0.5\% at low momentum to 1.0\%
at $200\gev$\,(natural units with $c=\hbar=1$ are used throughout this Letter).
The momentum scale is calibrated using samples of $\jpsi \rightarrow \mu^+ \mu^-$
and $\Bu \rightarrow \jpsi K^+$ decays collected concurrently with the data sample
used for this analysis~\cite{LHCb-PAPER-2011-035,LHCb-PAPER-2012-048,LHCb-PAPER-2013-011}.
The use of the large $\jpsi$ data sample allows to correct for variations of the momentum scale at the level of $10^{?4}$ or less that occur over time
whilst the use of the $\Bu \rightarrow \jpsi K^+$ decay allows the momentum scale to be determined as a function of the $K^+$ kinematics.
The procedure is validated  using samples of $\KS \rightarrow \pi^+ \pi^-$, $\psi(2S) \rightarrow \jpsi \pi^+ \pi^-$, $\psi(2S) \rightarrow
\mu^+ \mu^-$, other fully reconstructed $\bquark$-hadron and $\PUpsilon(nS), n=1,2,3$~decays. Based upon these studies the accuracy
of the procedure is evaluated to be $3 \times 10^{-4}$. 
Muons are identified by a
system composed of alternating layers of iron and multiwire
proportional chambers~\cite{LHCb-DP-2012-002}.
The online event selection is performed by a trigger~\cite{LHCb-DP-2012-004}, 
which consists of a hardware stage, based on information from the
calorimeter and muon systems, followed by a software stage, which applies a full event
reconstruction. 
The events used in this analysis are selected by a hardware
trigger that requires one or two muons with transverse momentum, \pt,  
larger than~$1.5 \gev$. 
At the software trigger stage, a pair of
oppositely charged muons with an invariant mass consistent with the known
$\jpsi$ mass~\cite{PDG2017} is required.
In Run\,1 the full event information for
selected events was stored.
To~keep the~rate within the available bandwidth it was
necessary to require $\pt(\jpsi) > 3 \gev$.
For Run\,2, a new data
taking scheme was introduced~\cite{Sciascia:2016dxb}
allowing real-time alignment to be
performed in the trigger~\cite{LHCb-PROC-2015-011}
that, together with an~increase in the online computing resources,
made possible
the full track reconstruction in
the online system~\cite{LHCb-DP-2016-001,Dziurda:2016oof}. Consequently, lower-level
information could be discarded, reducing the event size and allowing all
events selected at the hardware stage that contain a $\jpsi$ candidate 
to be stored without any $\pt$ requirement. 

Offline, $\jpsi$ candidates are combined with a pair of oppositely
charged muons to form $\chi_{c1,c2} \rightarrow \jpsi \mu^+ \mu^-$
candidates.  Several criteria are applied       
to reduce the background and      
maximize the sensitivity for the mass measurement.
Selected muon candidates are required to be within the range $2< \eta < 4.9$. 
Misreconstructed tracks are suppressed by the use of
a~neural network trained to discriminate between these and real
particles. 
Muon candidates are selected with a neural network
trained using simulated samples to discriminate muons from hadrons and electrons. 
Finally, to improve the mass resolution,
a kinematic fit is performed~\cite{Hulsbergen:2005pu}. 
In this fit the~mass of the~$\jpsi$~candidate is constrained to the known
mass of the~$\jpsi$~meson~\cite{PDG2017}
and the position of the $\chi_{c1,c2}$~candidate decay vertex is constrained
to be the same as that of the primary vertex.
The $\chi^2$ per degree of freedom of this fit is 
required to be less than four, which substantially reduces the
background while retaining almost all the signal
events. 

In the simulation, $pp$ collisions are generated using
\pythia~\cite{Sjostrand:2007gs}  with a specific
LHCb configuration \cite{LHCb-PROC-2010-056}.  For this study, signal
decays are generated  using \evtgen~\cite{Lange:2001uf} with decay
amplitudes that depend on the invariant dimuon mass, $m(\mu^+\mu^-)$, using the model described in
Ref.~\cite{Faessler:1999de}.
This model assumes that the decay proceeds via the emission of a virtual photon from
a~pointlike meson and is known to provide a good description of the corresponding dielectron
mode~\cite{Ablikim:2017kia}. 
Final-state radiation is accounted for using \photos~\cite{Golonka:2005pn}. 
The interaction of the generated particles with the detector, and its response,
are implemented using the \geant toolkit~\cite{Allison:2006ve, *Agostinelli:2002hh} 
as described in Ref.~\cite{LHCb-PROC-2011-006}.


The signal yields and parameters of the $\chi_{c1,c2}$ resonances are
determined with an extended unbinned maximum likelihood fit 
performed to the $\jpsi \mu^+
 \mu^-$ invariant mass distribution. 
In this fit, the $\chicone$ and $\chictwo$
signals are modelled by relativistic Breit-Wigner functions with
Blatt-Weisskopf form factors~\cite{Blatt:1952ije} with a meson radius
parameter of $3\gev^{-1}$. 
Jackson form factors~\cite{Jackson:1964zd}
are considered as an alternative to estimate the uncertainty associated
with this choice.    
The orbital angular momentum between the~$\jpsi$~meson and the~$\mu^+\mu^-$~pair
is assumed to be 0\,(1) for the $\chicone\,(\chictwo)$~cases. 

The relativistic Breit-Wigner functions are convolved with the detector
resolution. 
Three resolution models are found to describe the simulated data well: 
a double-Gaussian function, 
a double-sided Crystal Ball function~\cite{Skwarnicki:1986xj, LHCB-Paper-2011-013} 
and a symmetric variant of the Apollonios function~\cite{Santos:2013gra}. 
The double\nobreakdash-Gaussian function is used by the default model and the other functions are 
considered to estimate the systematic uncertainty. 
The parameters of the resolution model are determined by a 
simultaneous fit to the $\chicone$ and $\chictwo$ simulated
samples. 
All the parameters apart from the core resolution parameter, 
$\sigma$, are common between the two decay modes. 
For all the models in the simulation it is found that 
$\alpha \equiv \sigma^{\chictwo}/\sigma^{\chicone} = 1.13 \pm 0.01$. 
This is close to the value expected,  
$\alpha = 1.11$, from the assumption that the resolution scales with the square
root of the energy release. 

Combinatorial background
is modelled by a second\nobreakdash-order polynomial function. 
The total fit function consists of the sum of the background and the $\chicone$ and
$\chictwo$ signals. The free parameters are the yields of the two
signal components, 
the yield of the~background component, 
the two background shape parameters, 
the $\chicone$ and $\chictwo$ masses, 
$\sigma^{\chicone}$ and the natural width of the
$\chictwo$~resonance, $\Gamma(\chictwo)$. 
The other resolution parameters are fixed to the simulation values.  
Since the natural width of the $\chicone$ state \mbox{$\Gamma(\chicone)=0.84 \pm 0.04\mev$}~\cite{PDG2017}  
is less than the detector resolution\,(\mbox{$\sigma^{\chicone} =  1.41 \pm 0.01 \mev$}),
this study has limited sensitivity to its value. By applying Gaussian constraints on
the natural width of the~\chicone state\,(to the value from Ref.~\cite{PDG2017}) 
and $\alpha$\,(to the value found in the simulation) the $\chictwo$ width is determined 
in a data-driven way using the observed resolution for the $\chicone$ state.


The fit of this model to the full data sample is shown in
Fig.~\ref{fig:fit:final} and the resulting parameters of 
interest are summarized in Table~\ref{tab:fits:result}. 
The fitted value of $\sigma^{\chicone}$ is $1.51 \pm 0.04 \mev$, 
which agrees at the level of $5 \%$ with the value
found in the simulation.
Figure~\ref{fig:data:mumumass} shows the
$m(\mumu)$ mass distribution for selected candidates where the
background has been subtracted using the $\sPlot$ technique
\cite{Pivk:2004ty}. The data agree well with the model described in
Ref.~\cite{Faessler:1999de}.

\ifthenelse{\boolean{forprl}}{

\begin{figure}[tb]
  \resizebox{\columnwidth}{!}{\includegraphics*{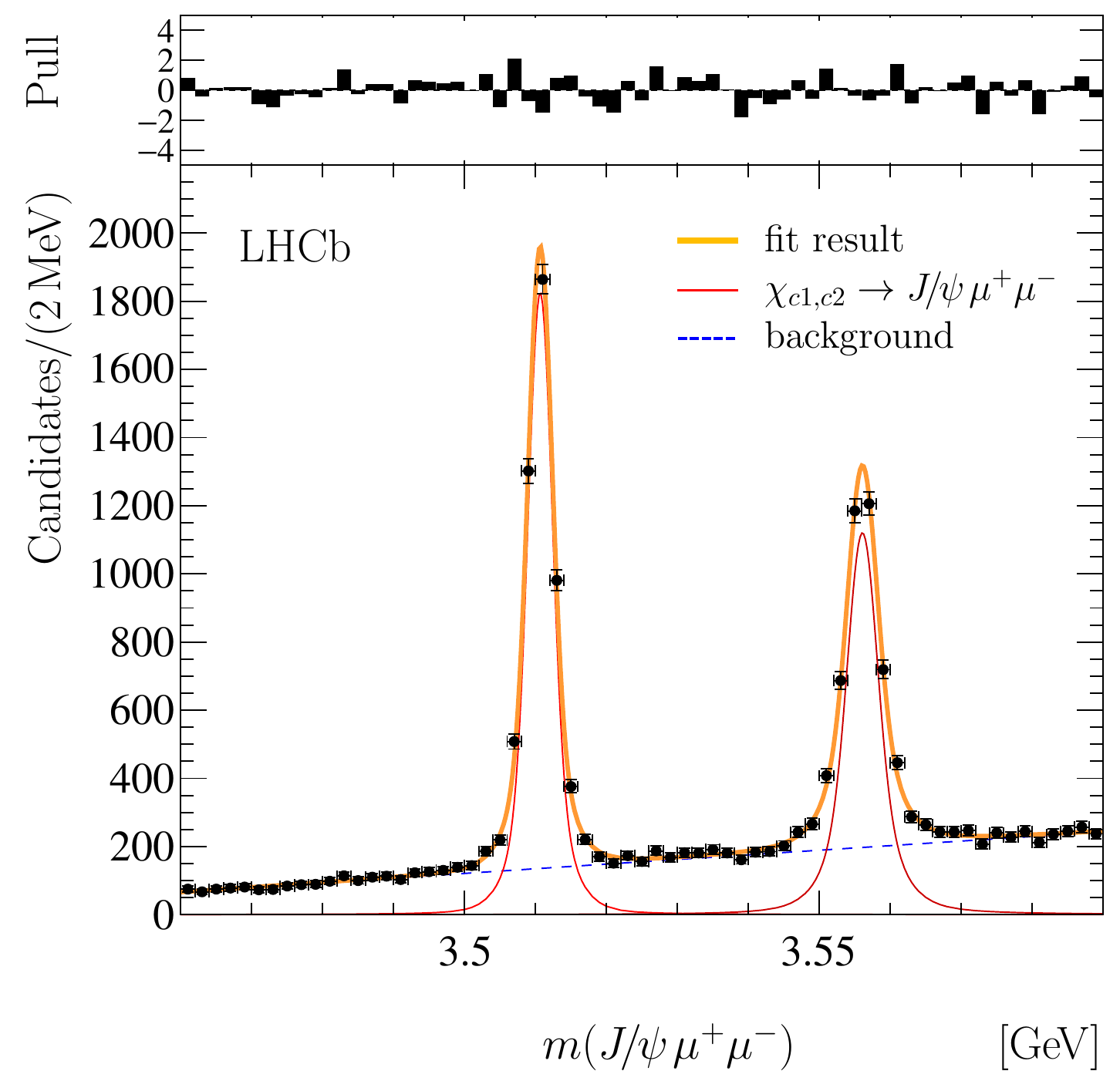}}
  \caption { \small
    Mass distribution for selected $\jpsi\mumu$~candidates.
    The~fit is shown in thick orange, the~$\chicone$
    and $\chictwo$ signal components are shown by the~thin red solid curve and
    the background component by the dashed blue curve.
  }
  \label{fig:fit:final}
\end{figure}

\begin{table}[tb]
  \centering
  \caption{ \small
    Signal yields and resonance parameters 
    from the nominal fit.
    No correction for final-state radiation is applied 
    to the mass measurements at this stage.
  }
\vspace*{2mm}
\begin{tabular*}{60mm}{@{\hspace{4mm}}ll@{\extracolsep{\fill}}c@{\hspace{4mm}}}
   \multicolumn{2}{c}{Fit parameter}     & Fitted value
  \\ \hline
    $N(\chicone)$     &  &   $\phantom{0.00}4\,755\pm81\phantom{0.0}$      
    \\
    $N(\chictwo)$     &  &   $\phantom{0.00}3\,969\pm96\phantom{0.0}$        
    \\
    $m(\chicone)$      &  $\left[\!\mev\right]$  &   $3\,510.66\pm0.04$
    \\
    $m(\chictwo)$      &  $\left[\!\mev\right]$  &   $3\,556.07\pm0.06$
    \\
    $\Gamma(\chictwo)$ &  $\left[\!\mev\right]$  &   $\phantom{0\,00}2.10\pm0.20$ 
\end{tabular*}
\label{tab:fits:result}
\end{table}
\begin{figure}[tb]
  \resizebox{\columnwidth}{!}{\includegraphics*{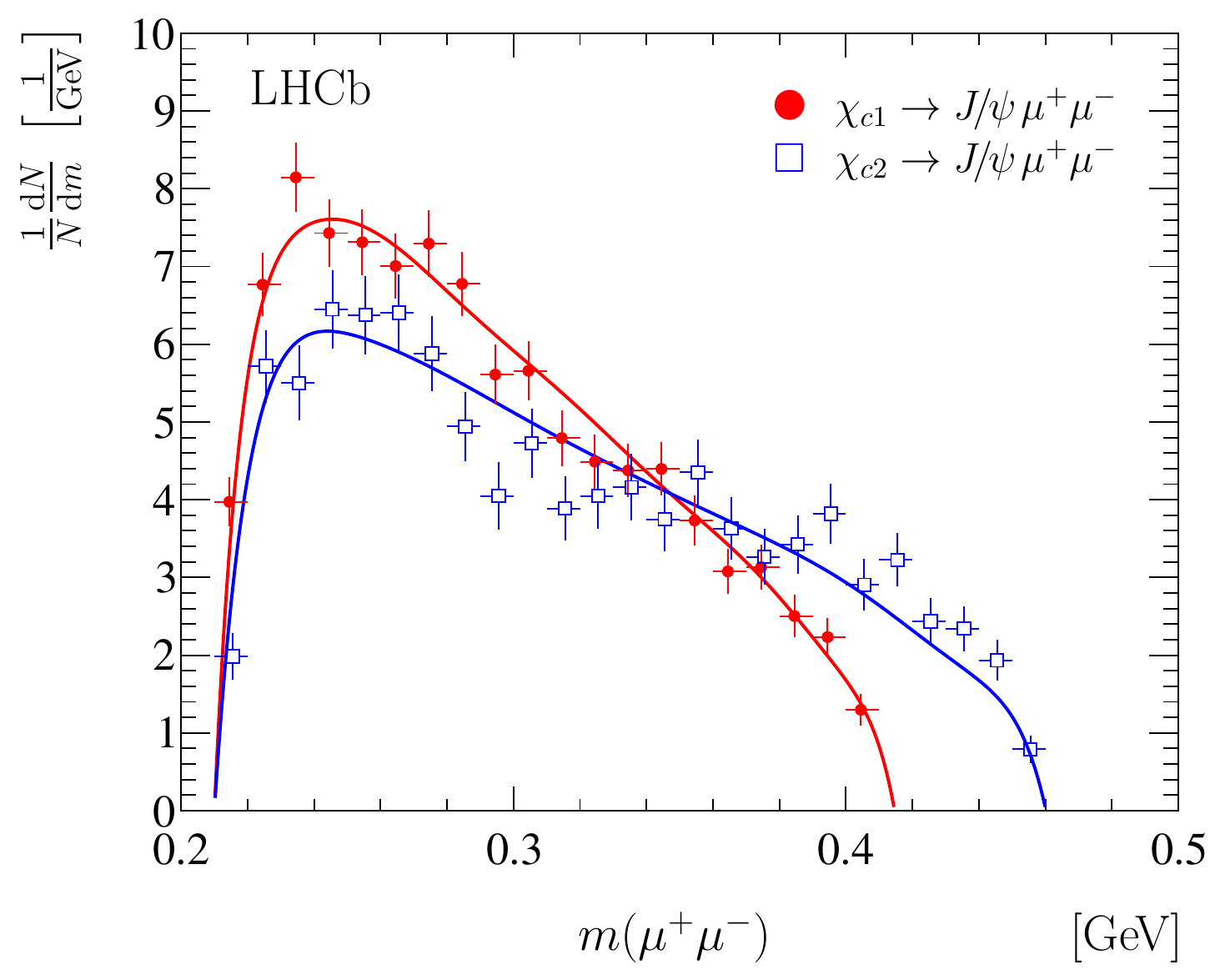}}
  \caption { \small
    Background-subtracted 
    $m(\mumu)$~distribution
    for
    $\mbox{\chicone\to\jpsi\mumu}$~\,(solid red  circles) 
    and 
    $\mbox{\chictwo\to\jpsi\mumu}$\,(open blue squares)~decays. The
    distributions are normalized to unit area. The~curves show the expected distribution from the simulation,
    which uses the model described in Ref.~\cite{Faessler:1999de}.  
  }\label{fig:data:mumumass}
\end{figure}
}{

\begin{figure}[t]
  \setlength{\unitlength}{1mm}
  \centering
  \begin{picture}(150,150)
    \put(  0, 0){
      \includegraphics*[width=150mm,height=150mm,%
      ]{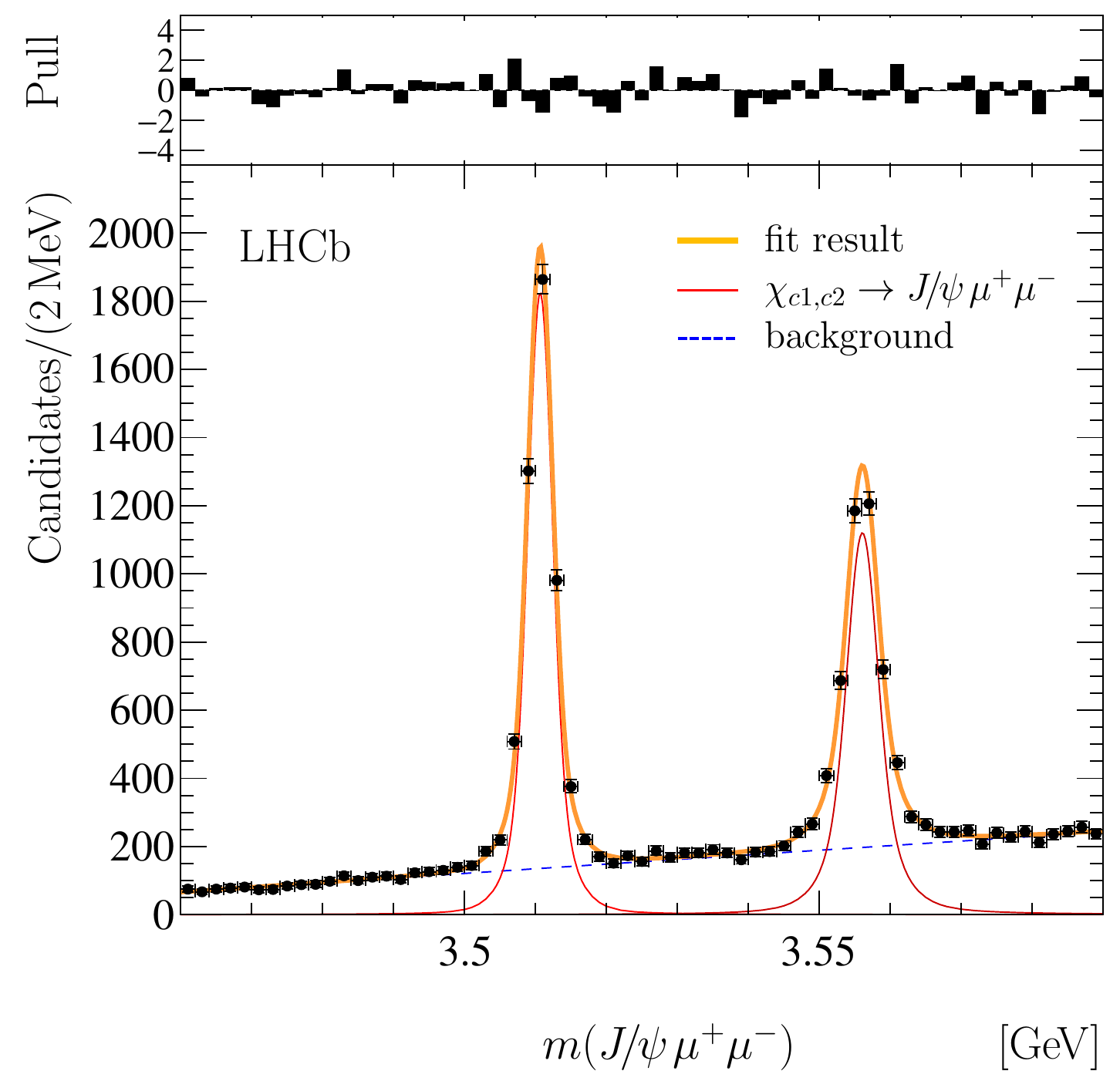}
    }
  \end{picture}
  \caption { \small
    Mass distribution for selected $\jpsi\mumu$~candidates.
    The~fit is shown in thick orange, the~$\chicone$
    and $\chictwo$ signal components are shown by the~thin red solid curve and
    the background component by the dashed blue curve.
  }
  \label{fig:fit:final}
\end{figure}

\begin{table}[t]
  \centering
  \caption{ \small
    Signal yields and resonance parameters 
    from the nominal fit.
    No correction for final-state radiation is applied 
    to the mass measurements at this stage.
  }
  \vspace*{2mm}
  \begin{tabular*}{75mm}{@{\hspace{4mm}}ll@{\extracolsep{\fill}}c@{\hspace{4mm}}}
    \multicolumn{2}{c}{Fit parameter}     & Fitted value
    \\ \hline
    $N(\chicone)$      &  &   $\phantom{0.00}4755\pm81\phantom{0.0}$      
    \\
    $N(\chictwo)$      &  &   $\phantom{0.00}3969\pm96\phantom{0.0}$        
    \\
    $m(\chicone)$      &  $\left[\!\mev\right]$  &   $3510.66\pm0.04$
    \\
    $m(\chictwo)$      &  $\left[\!\mev\right]$  &   $3556.07\pm0.06$
    \\
    $\Gamma(\chictwo)$ &  $\left[\!\mev\right]$  &   $\phantom{000}2.10\pm0.20$ 
\end{tabular*}
\label{tab:fits:result}
\end{table}

\begin{figure}[t]
  \setlength{\unitlength}{1mm}
  \centering
  \begin{picture}(150,120)
    \put(  0, 0){
      \includegraphics*[width=150mm,height=120mm,%
      ]{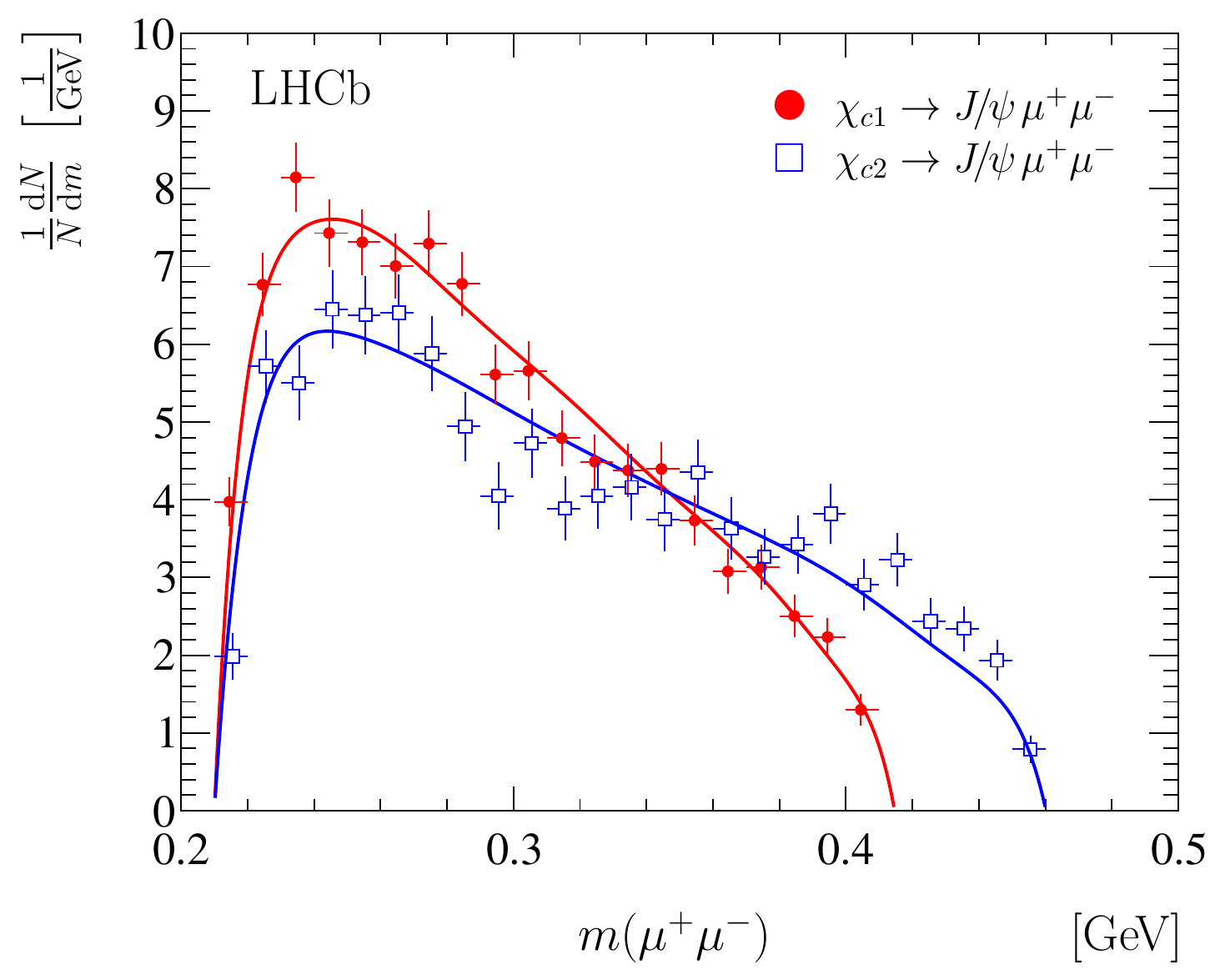}
    }
  \end{picture}
  \caption { \small
    Background-subtracted 
    $m(\mumu)$~distribution
    for
    $\mbox{\chicone\to\jpsi\mumu}$~\,(solid red  circles) 
    and 
    $\mbox{\chictwo\to\jpsi\mumu}$\,(open blue squares)~decays. The
    distributions are normalized to unit area. The~curves show the expected distribution from the simulation,
    which uses the model described in Ref.~\cite{Faessler:1999de}.  
  }\label{fig:data:mumumass}
\end{figure}

}

The dominant source of systematic uncertainty on the mass measurements
comes from the knowledge of the momentum scale. This is evaluated by
adjusting the momentum scale by the $3 \times 10^{-4}$
uncertainty on the calibration procedure and rerunning the mass fit. Uncertainties
of $88\kev$ and $102 \kev$ are assigned to the 
$\chicone$ and $\chictwo$ mass measurements, respectively. 
A further
uncertainty arises from the knowledge of the correction for energy
loss in the spectrometer, which is known with 10\,\%
accuracy~\cite{LHCb-DP-2014-002}. 
Based on the studies in Ref. \cite{LHCb-PAPER-2013-011}
a $20 \kev$ uncertainty is assigned. 

The distortion of the lineshape due to final-state radiation 
introduces a bias on the mass. 
This bias is evaluated using the simulation to be 
\mbox{$47 \pm 7 \kev\,(29\pm 10 \kev)$} for the 
\mbox{$\chicone\,(\chictwo)$} where the uncertainty is
statistical. 
The central values of the mass measurements are corrected accordingly and
the uncertainties are propagated.

Other uncertainties arise from the fit modelling and are studied using
a simplified simulation. Several variations of the relativistic
Breit-Wigner distribution are considered. Using Jackson form factors,
modifying  the meson radius parameter and varying the orbital angular 
momentum, the observed \mbox{$\chicone\,(\chictwo)$}~mass changes 
by at most $15\,(24)\kev$, which is assigned as a systematic
uncertainty. Similarly, fitting with a double-sided Crystal Ball or
Apollonios model, variations of $7 \kev$ and $2 \kev$ are seen
for the $\chicone$  and $\chictwo$  masses 
and assigned as systematic uncertainties. Finally, varying the order
of the polynomial background function results in a further
uncertainty of $2 \kev$.
The uncertainties due to the
momentum scale and energy loss correction largely cancel in the mass difference.
The assigned systematic uncertainties on the mass measurements are summarized in
Table~\ref{tab:MassSyst}.

\ifthenelse{\boolean{forprl}}{
\begin{table}[t]
\caption{\small Systematic uncertainties (in \!\kev) on the
  mass and mass difference measurements.}
\label{tab:MassSyst}
\vspace*{2mm}
\centering
\small
\begin{tabular*}{\columnwidth}{@{\hspace{1mm}}l@{\extracolsep{\fill}}ccc@{\hspace{1mm}}}
Source of uncertainty & $m(\chicone)$ & $m(\chictwo)$ &
$m(\chictwo) - m(\chicone)$ \\ \hline
Momentum scale                      & $88$             & $102$            & $18$        \\
Energy loss correction              & $20$             & $\phantom{0}20$  & ---         \\
Final-state radiation               & $\phantom{0}7$   & $\phantom{0}10$  & $12$            \\
Resonance shape                     & $15$             & $\phantom{0}24$  & $25$           \\
Background model                    & $<2\phantom{0}$  & $<2$             & $<2\phantom{0}$ \\
Resolution model                    & $\phantom{0}7$   & $\phantom{00}2$  & $\phantom{0}6$ \\
\hline
Sum in quadrature  & 92 & 107 & 34  \\
\end{tabular*}
\end{table}
}{
\begin{table}[t]
\centering
\caption{\small Systematic uncertainties on the
  mass and mass difference measurements.}
\label{tab:MassSyst}
\vspace*{2mm}
\begin{tabular*}{0.95\columnwidth}{@{\hspace{1mm}}l@{\extracolsep{\fill}}ccc@{\hspace{1mm}}}
Source of uncertainty & $m(\chicone)\,\left[\!\kev\right]$ & $m(\chictwo)\,\left[\!\kev\right]$ &
$m(\chictwo) - m(\chicone)\,\left[\!\kev\right]$ \\ \hline
Momentum scale                      & $88$             & $102$            & $18$        \\
Energy loss correction              & $20$             & $\phantom{0}20$  & ---         \\
Final-state radiation               & $\phantom{0}7$   & $\phantom{0}10$  & $12$            \\
Resonance shape                     & $15$             & $\phantom{0}24$  & $25$           \\
Background model                    & $<2\phantom{0}$  & $<2$             & $<2\phantom{0}$ \\
Resolution model                    & $\phantom{0}7$   & $\phantom{00}2$  & $\phantom{0}6$ \\
\hline
Sum in quadrature  & 92 & 107 & 34  \\
\end{tabular*}
\end{table}
}

The main uncertainty on the determination of the natural width of the
$\chictwo$ is due to the knowledge of the detector resolution. This is
accounted for in the statistical uncertainty since the resolution scale is
determined using the $\chicone$ signal in data. Similarly, the
uncertainty on the knowledge of the $\chicone$ width is propagated via
the Gaussian constraint in the mass fit. By running fits with and without
the constraint the latter is evaluated to be $40\kev$. Further uncertainties of
$10\kev$ and $20\kev$ arise from the assumed Breit-Wigner parameters
and resolution model, respectively. Other systematic uncertainties,
\eg\ due to the background model, are negligible.
The stability of the
results is studied by dividing the data into different running periods and also into
kinematic bins and repeating the fit.
None of these tests shows
evidence of a systematic bias.

In summary, the decays \mbox{$\chicone \rightarrow  \jpsi \mu^+ \mu^-$} and
\mbox{$\chictwo\rightarrow \jpsi \mu^+ \mu^-$} are observed and the mass
of the $\chicone$ meson together with the mass
and natural width of the $\chictwo$  are measured. The results for the
mass measurements are
\begin{align*}  
  m(\chicone)              & = 3510.71           \pm 0.04 \pm 0.09\mev, \\
  m(\chictwo)              & = 3556.10           \pm 0.06 \pm 0.11\mev, \\ 
  m(\chictwo)- m(\chicone) & = \phantom{00}45.39 \pm 0.07 \pm 0.03\mev, 
\end{align*}
where the first uncertainty is statistical and the  second is systematic.
The dominant systematic uncertainty is due to the knowledge of
the momentum scale and largely cancels in the mass difference.
It can be seen in Table~\ref{tab:summary} that the
measurements are in good agreement with
and have comparable precision to the best previous ones, 
made using $p\overline{p}$ annihilation at threshold by the
E760~\cite{Armstrong:1991yk} and E835 experiments~\cite{Andreotti:2005ts} 
at Fermilab.
They are considerably more precise
than the best measurement
based on the final\nobreakdash-state reconstruction~\cite{Ablikim:2005yd}.
It should be noted that the world average for the $\chicone$ mass has a
scale factor of 1.5 to account for the poor agreement between the results~\cite{PDG2017}.
The result for the $\chictwo$ natural width is
\begin{equation*}
  \Gamma(\chictwo) = 2.10 \pm  0.20\stat \pm 0.02\syst\mev\,. \\
\end{equation*}
It has similar precision to and is in good agreement with 
previous measurements \cite{PDG2017}.

\ifthenelse{\boolean{forprl}}{\begin{table}[t]
  \caption{\small LHCb measurements, compared to both previous
    measurements from Ref.~\cite{Armstrong:1991yk} and the current 
    world averages from Ref.~\cite{PDG2017}.
    The quoted uncertainties includes statistical and systematic uncertainties.}
\vspace*{2mm}
\centering
\begin{tabular*}{\columnwidth}{@{\hspace{1mm}}l@{\extracolsep{\fill}}ccc@{\hspace{1mm}}}
 Quantity &   LHCb   & Best previous &  \\ 
 $\,\left[\!\mev\right]$ & measurement & measurement   & \raisebox{1.5ex}[-1.5ex]{World average} \\
\hline
$m(\chicone) $    & $3510.71 \pm 0.10$ & $3510.72 \pm 0.05$
                                              & $3510.66 \pm 0.07$ \\
$m(\chictwo)$   & $3556.10 \pm 0.13$ & $3556.16 \pm 0.12$
                                              & $3556.20 \pm 0.09$ \\
$\Gamma(\chictwo)$ & $\phantom{000}2.10 \pm 0.20$ & $\phantom{000}1.92\pm 0.19$ & $\phantom{000} 1.93 \pm 0.11$ \\
\end{tabular*}
\label{tab:summary}
\end{table}

}{
\begin{table}[t]
  \caption{\small LHCb measurements compared to both previous
    measurements from Ref.~\cite{Armstrong:1991yk} and 
    the current world averages from Ref.~\cite{PDG2017}.
    The quoted uncertainties includes statistical and systematic uncertainties.}
  \vspace*{2mm}
  \centering
  \begin{tabular*}{0.80\textwidth}{@{\hspace{1mm}}l@{\extracolsep{\fill}}ccc@{\hspace{1mm}}}
    Quantity                &   LHCb      & Best previous &  \\ 
    $\,\left[\!\mev\right]$ & measurement & measurement   & \raisebox{1.5ex}[-1.5ex]{World average} \\
    \hline
    $m(\chicone) $     & $3510.71 \pm 0.10$ & $3510.72 \pm 0.05$           & $3510.66 \pm 0.07$ \\
    $m(\chictwo)$      & $3556.10 \pm 0.13$ & $3556.16 \pm 0.12$           & $3556.20 \pm 0.09$ \\
    $\Gamma(\chictwo)$ & $\phantom{000}2.10 \pm 0.20$    & $\phantom{000} 1.92\pm 0.19$ & $\phantom{000} 1.93 \pm 0.11$ \\
  \end{tabular*}
  \label{tab:summary}
\end{table}

}

The observations presented here open up a new avenue for hadron
spectroscopy at the LHC. These decay modes can be used 
to measure the production 
of  $\chicone$ and $\chictwo$~states 
with a similar precision to the converted photon study presented 
in Ref.~\cite{LHCb-PAPER-2013-028}. Importantly, it will be possible to
extend measurements down to very low $\pt(\chi_{c1,c2})$ probing further QCD
predictions \cite{Ma:2010vd,Likhoded:2015qyl,Boer:2012bt}. 
In addition, measurements of the transition form
factors~\cite{Landsberg:1986fd} will provide inputs on the interaction
between charmonium states and the electromagnetic field. 
With larger data samples, 
studies of the Dalitz decays of other heavy-flavour
states will become possible. 
For example, measurement of the transition form factor of the $X(3872)$ via its Dalitz
decay may help elucidate the nature of this enigmatic state~\cite{Ablikim:2017kia}.

%
%
\section*{Acknowledgements}
\noindent 
We thank  Jielei Zhang for useful
discussions concerning Dalitz decays. 
We~express our gratitude to our colleagues in the CERN
accelerator departments for the excellent performance of the LHC. We
thank the technical and administrative staff at the LHCb
institutes. 
We~acknowledge support from CERN and from the national
agencies: 
CAPES, CNPq, FAPERJ and FINEP\,(Brazil); 
MOST and NSFC\,(China); 
CNRS/IN2P3\,(France); 
BMBF, DFG and MPG\,(Germany); 
INFN\,(Italy); 
NWO\,(The~Netherlands); 
MNiSW and NCN\,(Poland); 
MEN/IFA\,(Romania); 
MinES and FASO\,(Russia); 
MinECo\,(Spain); 
SNSF and SER\,(Switzerland); 
NASU\,(Ukraine); 
STFC\,(United Kingdom); 
NSF\,(USA).  
We~acknowledge the computing resources that are provided by CERN, 
IN2P3\,(France), 
KIT and DESY\,(Germany), 
INFN\,(Italy), 
SURF\,(The~Netherlands),
PIC\,(Spain), 
GridPP\,(United Kingdom), 
RRCKI and Yandex LLC\,(Russia), 
CSCS\,(Switzerland), 
IFIN\nobreakdash-HH\,(Romania), 
CBPF\,(Brazil),
PL\nobreakdash-GRID\,(Poland) 
and OSC\,(USA). 
We~are indebted to the communities
behind the multiple open-source software packages on which we depend.
Individual groups or members have received support from 
AvH Foundation\,(Germany), 
EPLANET, Marie Sk\l{}odowska\nobreakdash-Curie Actions and ERC\,(European Union), 
ANR, Labex P2IO, ENIGMASS and OCEVU, and R\'{e}gion Auvergne\nobreakdash-Rh\^{o}ne\nobreakdash-Alpes\,(France), 
RFBR and Yandex LLC\,(Russia), 
GVA, XuntaGal and GENCAT\,(Spain), 
Herchel Smith Fund, the Royal Society,
the English\nobreakdash-Speaking Union and the Leverhulme Trust\,(United Kingdom)



\clearpage
\addcontentsline{toc}{section}{References}
\setboolean{inbibliography}{true}
\bibliographystyle{LHCb}
\bibliography{main,local,LHCb-PAPER,LHCb-CONF,LHCb-DP,LHCb-TDR}

\newpage


\centerline{\large\bf LHCb collaboration}
\begin{flushleft}
\small
R.~Aaij$^{40}$,
B.~Adeva$^{39}$,
M.~Adinolfi$^{48}$,
Z.~Ajaltouni$^{5}$,
S.~Akar$^{59}$,
J.~Albrecht$^{10}$,
F.~Alessio$^{40}$,
M.~Alexander$^{53}$,
A.~Alfonso~Albero$^{38}$,
S.~Ali$^{43}$
% ...

{\footnotesize \it
% ...
$ ^{5}$Clermont Universit{\'e}, Universit{\'e} Blaise Pascal, CNRS/IN2P3, LPC, Clermont-Ferrand, France\\
% ...
$ ^{10}$Fakult{\"a}t Physik, Technische Universit{\"a}t Dortmund, Dortmund, Germany\\
% ...
$ ^{38}$ICCUB, Universitat de Barcelona, Barcelona, Spain\\
$ ^{39}$Universidad de Santiago de Compostela, Santiago de Compostela, Spain\\
$ ^{40}$European Organization for Nuclear Research (CERN), Geneva, Switzerland\\
% ...
$ ^{43}$Nikhef National Institute for Subatomic Physics, Amsterdam, The Netherlands\\
% ...
$ ^{48}$H.H. Wills Physics Laboratory, University of Bristol, Bristol, United Kingdom\\
% ...
$ ^{53}$School of Physics and Astronomy, University of Glasgow, Glasgow, United Kingdom\\
% ...
$ ^{59}$University of Cincinnati, Cincinnati, OH, United States\\
% ...
}
\end{flushleft}

\end{document}